\documentclass{PoS}

\title{Dark Matter in the Milky Way}

\ShortTitle{Dark Matter in the Milky Way}

\author{\speaker{Benoit Famaey}\\
        Observatoire astronomique de Strasbourg, Universit\'e de Strasbourg, CNRS, UMR 7550 \\ 11 rue de l'Universit\'e, F-67000 Strasbourg, France\\
        E-mail: \email{benoit.famaey@astro.unistra.fr}}

\abstract{We review most dynamical constraints on the gravitational field of spiral galaxies in general, and of the Milky Way in particular. Such constraints are of prime importance for determining the charateristics of the putative dark matter haloes of galaxies. For the Milky Way, we review observational constraints in the inner parts (cored or cusped dark matter distribution, maximum disk or not), in the solar neighbourhood (local dark matter density) and in the outer parts (virial mass and triaxial shape of the dark matter halo). We also point out various caveats, systematic effects, and large current uncertainties. Many fundamental parameters such as the local circular velocity are poorly known, evidence for triaxiality of the dark halo is shaky, and different estimates of the virial mass as well as of the local dark matter density vary by at least a factor of two. We however argue that the current best-fit value for the local dark matter density, which should be used as a benchmark for direct dark matter detection searches, is of the order of 0.5 ${\rm GeV} \, {\rm cm}^{-3}$. We also explain why alternatives to particle dark matter on galactic scales should still be very seriously considered.}

\FullConference{Frontiers of Fundamental Physics 14 - FFP14,\\
                15-18 July 2014\\
                Aix Marseille University (AMU) Saint-Charles Campus, Marseille }
		
\begin{document}

\section{Introduction}
\vspace{-3mm}
Inconsistencies between the rotation curves of spiral galaxies and the ones computed from their visible material in Newtonian gravity dates back to the late 1930's \cite{Babcock}, but became widely accepted as the ``missing mass problem'' only in the late 1970's to early 1980s \cite{Bosma,Rubin}, notably with the advent of radioastronomy allowing to probe the rotation curves well beyond the optical radius of galactic disks. Since then, various lines of evidence for this missing mass, or at least for the presence of a new degree of freedom in the Lagrangian of nature, have accumulated on scales ranging from the largest cosmological scales down to galactic scales. Perhaps the most convincing evidence that this new degree of freedom must behave like a dissipationless dust fluid on the largest scales is the relative height of the second and third peak in the angular power spectrum of the Cosmic Microwave Background \cite{Planck}: in the absence of a net forcing term decoupled from the baryons-photon fluid, the fluctuations should be diffusion-damped at decoupling, and more so on smaller scales. But the second and third peak are observed to have almost the same height, which thus provides evidence for a new, non-baryonic, degree of freedom. Obviously, the simplest framework for such a new degree of freedom decoupled from the baryons-photon plasma in the early Universe is a fluid of stable elementary non-baryonic particles interacting with each other and with baryons almost entirely through gravity, without any additional fundamental property encoded in their Lagrangian. However, it is important to keep in mind that there is currently no experimental evidence on galaxy scales that the new degree of freedom is indeed made of such ``particle dark matter'': indeed, various alternatives can reproduce the Cosmic Microwave Background whilst predicting that no dark matter particles should be detected in galaxies where the new degree of freedom could in fact be of a different nature, effectively modifying gravity at these scales rather than simply providing additional mass \cite{FMcG,Khoury,Blanchet}. Here, we review current constraints on the dark matter distribution in spiral galaxies in general, and in the Milky Way in particular, and we highlight the current large uncertainties which will hopefully be alleviated thanks to the ucoming data of the Gaia mission.

\vspace{-3mm}
\section{Dark matter in spiral galaxies}
\vspace{-3mm}
\subsection{Rotation curves}
\vspace{-1mm}
Within the framework of particle dark matter, the distribution of dark matter in galaxies in general, and in spiral galaxies in particular, has by now been a matter of debate for decades. Numerical simulations of the collapse of pure dark matter halos in the currently favored cosmological model lead to a density distribution as a function of radius, $\rho$, which is well fit by a smooth function asymptoting to a central cusp with slope ${\rm dln}\rho/{\rm dln}r \propto -r^{(1/n)} \sim -1$ in the central parts \cite{Navarro} (with $n \sim 6$ for a Milky Way-sized halo, meaning that the slope is still $-1$ at 200~pc from the center and asymptotes to 0 only at the very center). However, it appears that this does not correspond to what is deduced from actual rotation curves, and is known as the core-cusp problem.

In large spiral galaxies whose baryonic contribution is dominated by the stellar component, the actual dark matter distribution deduced from observations heavily depends on the stellar mass-to-light ratio ($M/L$). While this ratio is expected to be more constant as a function of colour in the near-infrared than in the optical part of the spectrum, recent stellar population synthesis models diverge by as much as a factor of three on the actual value of $M/L$ (and hence on the stellar mass of galaxies) in the $K$-band, ranging from $0.2 \, M_\odot/L_\odot$ to $0.7 \, M_\odot/L_\odot$. Moreover, the same model often yields a different stellar mass when applied to different bands. Recently, adopting the $V$-band as a reference point grounded in Milky Way stellar counts (see Sect.~3), it was shown \cite{McGSch} how models should be corrected to yield self-consistent galactic stellar masses, yielding $M/L_K = 0.6 \, M_\odot/L_\odot$. This value is grossly consistent with the ``maximum disk'' hypothesis, i.e. that the disk contributes maximally to the rotation curve. This is also backed by the measurement of the corotation radius in barred galaxies being not far beyond the bar end \cite{Sellwood}, which would not be possible if a lot of angular momentum had been transferred to a dominant dark matter halo. 

With such values of the stellar $M/L$, most rotation curves cannot be fit by a cusped profile such as predicted by the simulations, but rather by a cored profile (or at least by values of $n$ in the profile hereabove that effectively lead to a constant density core of $\sim {\rm kpc}$ size). Note that in the faintest, gas-dominated galaxies, this issue of the stellar $M/L$ is less crucial, as the gas mass is dominating, and a cored dark mater distribution is also most often required. The state-of-the-art solution to this problem is to rely on baryonic physics by enforcing strong supernovae outflows that move large amounts of low-angular-momentum gas from the central parts and that pull on the central dark matter concentration to create a core; but this is still a relatively fine-tuned process, which fails to account for cored profiles in the faintest galaxies (where the core is the least dependent on stellar population synthesis models), and also fails to produce their observed baryon fractions ([stars+gas]/total). With the self-consistent value of the stellar $M/L_K=0.6$, it is also interesting to note that a unique relation between the baryonic mass of galaxies and their asymptotic circular velocity (related to their dark matter content), the so-called baryonic Tully-Fisher relation, is valid over more than five decades in mass with essentially zero scatter. This exact relation is valid for faint gas-dominated dwarf galaxies \cite{McGBTF} as well as for tidal dwarf galaxies \cite{Gentiletdg} that are supposedly devoid of dark matter, thereby posing a huge fine-tuning challenge to our understanding of galaxy formation within the standard cosmological model \cite{illustris}. More generally, there appears to exist a one-to-one relationship between the distribution of baryons (more precisely their surface density) and the gravitational field in spirals \cite{FMcG,Milgrom}, which is hard to explain from chaotic and haphazard behaviors expected in galaxy formation, including baryonic feedback and different assembly histories: this is the main motivation for still trying out alternatives to particle dark matter on small scales \cite{Khoury,Blanchet}.

\vspace{-3mm}
\subsection{Stellar velocity dispersions}
\vspace{-1mm}
As explained hereabove, one of the main points of contention in determining the dark matter distribution in large stellar-dominated spirals is their stellar $M/L$ ratio. We have seen that, in order for stellar population synthesis models to yield the same galactic stellar mass when applied to different bands, the value $M/L_K = 0.6 \, M_\odot/L_\odot$ should be adopted \cite{McGSch}. Nevertheless, an alternative and independent way to measure the stellar $M/L$ is to measure it dynamically, from the vertical velocity dispersions in close to face-on galaxies. This is what the DiskMass survey \cite{Swaters} was designed for, and it measured surprisingly low vertical velocity dispersions for galaxies of relatively large scale-lengths, hence implying relatively low-mass disks. But to actually derive the dynamical surface density from the measured vertical velocity dispersions, what would additionally be required is a knowledge of the scale-heights of these galaxies, which is of course not measurable for such near-to-face-on galaxies. The only feasible option is then to set the disk scale-heights to be similar to those of edge-on galaxies with similar scale-lengths. With such an assumption, the DiskMass survey managed to derive surprisingly low $K$-band mass-to-light ratios \cite{Swaters}, i.e. $M/L_K = 0.24 \pm 0.05 \, M_\odot/L_\odot$ in blatant contradiction with \cite{McGSch}. If this dynamically measured low value of the stellar $M/L$ is the true one, it is a game-changer for all that was said hereabove: (i) the baryonic Tully-Fisher relation would no longer be a unique relation over five decades in mass, but would display two different branches with different zero-points, one for small gas-dominated and one for large stellar-dominated galaxies; (ii) cuspy profiles would be in perfect accordance with the dark matter distribution in the central parts of most stellar dominated spiral galaxies (and now even too concentrated w.r.t. simulations), but would paradoxically remain problematic in the faintest gas-dominated galaxies; (iii) most disks would be largely submaximal. It will thus be of the utmost importance to confirm or refute the DiskMass results in coming years, especially to assess whether the velocity dispersions that have been measured via integrated light in face-on galaxies really correspond to the same stellar populations for which the scale-heights have been measured in edge-on galaxies, and really are representative of dynamically relaxed populations. At present, the DiskMass results pose a number of problems for other observations: such submaximal stellar disks appear to be in contradiction with, e.g., the measurement of the corotation radius in barred galaxies being not far beyond the bar-end, and with many other measurements in our own Milky Way galaxy, as we will see hereafter.

\vspace{-3mm}
\section{Dark matter in our Galaxy}
\vspace{-3mm}
\subsection{Rotation curve}
\vspace{-1mm}
Paradoxically, our position inside the disk of the Milky Way makes it difficult to measure its outer rotation curve with a similar precision as in external galaxies. This indeed requires to know the precise distance of tracers, and to dynamically model them by taking into account an asymmetric drift correction and possible effects of non-axisymmetries (see Sect.~3.2). For the inner rotation curve, the situation is better as we can make use of the tangent point method, but it still requires to know our distance from the Galactic center $R_0$, the local circular velocity at the Sun's position $V_{c0}$ and the peculiar velocity of the Sun with respect to this circular velocity. Estimates of $R_0$ can vary, at the very extremes, from $\sim$6.5~kpc to $\sim$9.5~kpc, while associated measurements of the local circular velocity vary from $\sim$180~km/s to $\sim$300~km/s, all these values being heavily model-dependent \cite{McMilB}. These values are degenerate with the peculiar motion of the Sun and in particular with its azimuthal component (through the proper motion of SgrA$^*$ of 30.24~km/s/kpc), which is currently unknown within a factor of five, lying in the interval between 5~km/s and 30~km/s.

Regarding the point of contention of stellar mass-to-light ratios, star counts in the solar cylinder in the well-measured $V$-band yield a stellar $M/L_V = 1.5 \pm 0.2 \, M_\odot/L_\odot$, for a color index $B-V=0.58$ \cite{Flynn} which is in line with the self-consistent models yielding $M/L_K = 0.6 \, M_\odot/L_\odot$ \cite{McGSch}. If the local circular velocity is in the lower end of current estimates ($<230$~ km/s), this would a priori mean that the Milky Way has a maximum disk, provided it does not have too large a disk scale-length, in apparent contradiction with the DiskMass results, and it is backed by the distribution of microlensing events and by dynamical measurements of the surface mass density (see Sect.~3.3).

\vspace{-3mm}
\subsection{Effects of baryonic non-axisymmetries}
\vspace{-1mm}
The Milky Way is known to host spiral arms as well as a central bar: these non-axisymmetric features can provide important insights on the inner dark matter distribution of the Galaxy, notably through their effects on gaseous motions in the $l-v$ diagrams. It was shown \cite{BEG} that non-circular motions in the central parts of the Galaxy could be reproduced by modelling the effects of the central bar and heavy spiral arms (with a large amplitude in mass) in a maximum disk model backed by the microlensing optical depth, leaving no room to trade mass from the stellar disk to the dark matter halo, and strongly favouring a cored dark matter halo with a logarithmic potential of core radius $r_0=10.7$~kpc and asymptotic velocity $v_\infty=220$~km/s (with substantial freedom in the exact values of these parameters). In general, modelling stellar dynamics in the Milky Way whilst neglecting the effects of these non-axisymmetric features can be dangerous since it is not clear that assuming axisymmetry and dynamical equilibrium to fit a benchmark model does not bias the results by forcing this benchmark axisymmetric model to fit non-axisymmetric features in the observations. For instance, we showed that large and small-scale velocity gradients, including in the vertical direction, can be generated by both the bar and spirals \cite{Siebert, Williams, Faure, Bovy}. This should be kept in mind regarding the recent dynamical models presented hereafter.

\vspace{-3mm}
\subsection{Stellar dynamical models}
\vspace{-1mm}
With current large spectroscopic and upcoming astrometric surveys, one way forward to determine the dark matter distribution is to construct equilibrium dynamical models, using Jeans theorem constraining the phase-space distribution function to depend only on three isolating integrals of motion. In principle, one can iterate the fits with different Galactic potentials until the best-fitting potential is found, giving access to the underlying mass distribution. This is the philosophy we followed in \cite{Bienayme}, where a parametrized separable potential was used together with a fixed distribution depending on three isolating integrals of the motion, and was fit to the kinematics of 4600 red clump giants from the RAVE survey in a cylinder of 500~pc radius around the Sun. This allowed us to demonstrate that the local dark matter density is of the order of $\rho_{DM}=0.5 {\rm GeV} \, {\rm cm}^{-3}$, which should be used as a benchmark for local direct dark matter detection searches instead of the usually lower values used (different estimates from other studies vary by about a factor of two), nevertheless keeping in mind the caveat of the possible effects of non-axisymmetries mentioned above. In a similar spirit, \cite{BR13} made the first direct dynamical measurement of the Milky Way disk surface density profile for galactocentric distances between 4~kpc and 9~kpc, thanks to the dynamical modelling of various mono-abundance populations among 16000 SEGUE G-dwarfs. Interestingly, the fitted potential parameters for different mono-abundance populations were inconsistent with each other, a problem which was circumvented by using the dynamical surface density of each fitted potential at the radius where it was best constrained by its respective mono-abundance population. They concluded that the mass-weighted Galactic disk scale-length is short, $R_d=2.15 \pm 0.14 \,$kpc, and that the Milky Way disk is maximal, in contradiction with what is expected from the results of the DiskMass study \cite{Swaters}. Interestingly, this short scale-length of the vertical force is also in tension with the scale-length expected to be measured in theories modifying gravity at the classical level in galaxies \cite{FMcG}.

\vspace{-3mm}
\subsection{Virial mass}
\vspace{-1mm}
The virial mass of a galaxy is usually defined as the mass $M_{200}$ enclosed in the radius $R_{200}$ within which the mean overdensity is 200 times the critical density for closure in the Universe. A sample of high-velocity stars from RAVE \cite{Smith} made it possible to estimate the local escape speed from the Galaxy at the solar position, which for an adiabatically contracted NFW halo yields $R_{200}=305 \,$kpc and $M_{200}=1.4 \times 10^{12} \, M_\odot$. More realistically, defining the escape speed not as the speed which allows a star to reach infinity, but rather to reach a cutoff radius motivated by simulations, typically leads to slightly higher virial masses for a given escape speed, e.g. $M_{200}=1.6 \times 10^{12} \, M_\odot$ which is the current standard value \cite{Piffl}, in accordance with the measured velocity of the Leo~I galaxy at 260~kpc. This nevertheless seems in tension with the small velocity dispersion $\sigma \simeq 50 \,$km/s of distant halo stars between 100 and 150~kpc \cite{Deason}, and with mass estimators based on Jeans equations applied to halo stars or to satellite galaxies (excluding Leo~I), typically leading to virial masses $M_{200} \leq 10^{12} \, M_\odot$, keeping in mind caveats such as the granularity of the stellar halo and the anisotropy parameter which could lead to variations of more than a factor of three in the virial mass estimated in that way. Such a skinny Milky Way was also argued to be favoured by models of the Sagittarius stellar stream \cite{Gibbons}, but estimating the Milky Way mass in this way is still heavily underconstrained, as the shape of halo is also playing an esssential role.

\vspace{-3mm}
\subsection{Triaxial halo shape}
\vspace{-1mm}
A last important point about the distribution of dark matter around the Milky Way is whether it is triaxial at large radii from the center. In principle, stellar streams from dissolved galaxy satellites accreted onto the Galaxy are the best current probe to measure this possible triaxiality of the dark halo. Recent analyses of the Sagittarius stream have claimed that the kinematics and positions of its M giant stars constrain the halo to be triaxial and extremely flattened, being essentially an oblate ellipsoid oriented perpendicular to the disk. However, there is a strong degeneracy in this fitting procedure between the triaxiality and the halo density profile, as some perfectly spherical configurations fit the stream as well as the triaxial solution \cite{Ibatasag}. Hence, current evidence for the triaxiality of the Milky Way dark matter halo at large distances from the center is still not robust.

\vspace{-3mm}
\section{Conclusions}
\vspace{-1mm}
In summary, our current knowledge of the distribution of putative dark matter in spiral galaxies in general, and in the Milky Way in particular, is still scarce. Forthcoming data from the Gaia mission will hopefully help improve this state of affairs. There are currently conflicting results on the maximality of galactic disks, and on the cored or cusped distribution of dark matter in the central parts of external spiral galaxies, but current Milky Way data indicate that its halo is most likely cored and its disk maximal. Different estimates of the virial mass and of the local dark matter density vary by at least a factor of two, but the most solid estimates are on the high end in both cases, $\rho_{DM}=0.5 {\rm GeV} \, {\rm cm}^{-3}$ and $M_{200}=1.6 \times 10^{12} \, M_\odot$. However, until dark matter particles are detected in the lab, it is also heatlhy to remember that there are hints that the dark sector might be less straightforward than stable elementary non-baryonic particles interacting with each other and with baryons almost entirely through gravity. Such hints are for instance the precise scaling relations with essentially zero scatter and more generally the one-to-one relationship between the distribution of baryons (more precisely their surface density) and the gravitational field. The other main reason to still try alternatives is the failure of models of galaxy formation within the standard cosmological model to account for positional and kinematic correlations of galaxy satellites around their hosts, both in the Local Group \cite{Kroupa,Pawlowski} and more generally in the low-redshift Universe \cite{Neil,Rod}. The future of this field of research might thus still be full of exciting surprises for astrophysicists, cosmologists, and theoretical physicists.

\end{document}